\documentclass[10pt,conference]{IEEEtran}
\usepackage{times}
\usepackage{cite}
\usepackage{mdwmath}
\usepackage[english]{babel}
\usepackage{epsfig}
\usepackage{easymat}
\usepackage{amsfonts}
\usepackage{bm}
\usepackage{amsbsy}
\usepackage{amsmath}
\usepackage{float}
\usepackage{cite}
\usepackage{graphicx}
\usepackage{balance}
\usepackage{subfigure}
\usepackage{latexsym}
\usepackage{amssymb}
\usepackage{txfonts}
\usepackage{wasysym}
\usepackage{mathbbol}
\usepackage{multirow}
\usepackage{stfloats}
\usepackage{epstopdf}
\newtheorem{theorem}{Theorem}

\begin{document}
%%%%%%%%%%%%%%%%%%%%%%%%%%%%%%%%%%%%%%%%%%%%%%%%%%%%%%%%%%%%%%%%%%%%%%%%%%%%%%%%%%%%%%%%%%%%%%%%%%%%%%%%%%%%%%%%%%%%%%%%
%%%%%%%%%%%%%%%%%%%%%%%%%%%%%%%%%%%%%%%%%%%%%%%%%%%%%%%%%%%%%%%%%%%%%%%%%%%%%%%%%%%%%%%%%%%%%%%%%%%%%%%%%%%%%%%%%%%%%%%%
%%%%%% TITLE %%%%%%
%%%%%%%%%%%%%%%%%%%%%%%%%%%%%%%%%%%%%%%%%%%%%%%%%%%%%%%%%%%%%%%%%%%%%%%%%%%%%%%%%%%%%%%%%%%%%%%%%%%%%%%%%%%%%%%%%%%%%%%%
%%%%%%%%%%%%%%%%%%%%%%%%%%%%%%%%%%%%%%%%%%%%%%%%%%%%%%%%%%%%%%%%%%%%%%%%%%%%%%%%%%%%%%%%%%%%%%%%%%%%%%%%%%%%%%%%%%%%%%%%%%%%%%%%%
%%%%%%%%%%%%%%%%%%%%%%%%%%%%%%%%%%%%%%%%%%%%%%%%%%%%%%%%%%%%%%%%%%%%%%%%%%%%%%%%%%%%%%%%%%%%%%%%%%%%%%%%%%%%%%%%%%%%%%%%%%%%%%%%%
\renewcommand{\textfraction}{0}
\title{On the Low SNR Capacity of Maximum Ratio Combining over Rician Fading Channels with Full Channel State Information}
\author{ Fatma Benkhelifa*, Student, Zouheir Rezki, IEEE member, and Mohamed-Slim Alouini, IEEE fellow \\
\small *Computer, Electrical and Mathematical Science and Engineering (CEMSE) Division\\
\small King Abdullah University of Science and Technology (KAUST)
\\
\small Thuwal, Makkah Province, Saudi Arabia \\
\small \{fatma.benkhelifa,zouheir.rezki,slim.alouini\}@kaust.edu.sa}
\date{}
\maketitle \thispagestyle{empty}
%%%%%%%%%%%%%%%%%%%%%%%%%%%%%%%%%%%%%%%%%%%%%%%%%%%%%%%%%%%%%%%%%%%%%%%%%%%%%%%%%%%%%%%%%%%%%%%%%%%%%%%%%%%%%%%%%%%%%%%%
%%%%%%%%%%%%%%%%%%%%%%%%%%%%%%%%%%%%%%%%%%%%%%%%%%%%%%%%%%%%%%%%%%%%%%%%%%%%%%%%%%%%%%%%%%%%%%%%%%%%%%%%%%%%%%%%%%%%%%%%
%%%%%% ABSTRACT %%%%%%
%%%%%%%%%%%%%%%%%%%%%%%%%%%%%%%%%%%%%%%%%%%%%%%%%%%%%%%%%%%%%%%%%%%%%%%%%%%%%%%%%%%%%%%%%%%%%%%%%%%%%%%%%%%%%%%%%%%%%%%%
%%%%%%%%%%%%%%%%%%%%%%%%%%%%%%%%%%%%%%%%%%%%%%%%%%%%%%%%%%%%%%%%%%%%%%%%%%%%%%%%%%%%%%%%%%%%%%%%%%%%%%%%%%%%%%%%%%%%%%%%%%%%%%%%%
%%%%%%%%%%%%%%%%%%%%%%%%%%%%%%%%%%%%%%%%%%%%%%%%%%%%%%%%%%%%%%%%%%%%%%%%%%%%%%%%%%%%%%%%%%%%%%%%%%%%%%%%%%%%%%%%%%%%%%%%%%%%%%%%%
\begin{abstract}
{\ In this letter, we study the ergodic capacity of a maximum ratio combining (MRC) Rician fading channel with full channel state information (CSI) at the transmitter and at the receiver. We focus on the low Signal-to-Noise Ratio (SNR) regime and we show that the capacity scales as $\frac{L \Omega}{K+L} \text{SNR}\times \log(\frac{1}{\text{SNR}})$, where $\Omega$ is the expected channel gain per branch, $K$ is the Rician fading factor, and $L$ is the number of diversity branches. We show that one-bit CSI feedback at the transmitter is enough to achieve this capacity using an on-off power control scheme. Our framework can be seen as a generalization of recently established results regarding the fading-channels capacity characterization in the low-SNR regime.\par}
\end{abstract}
%%%%%%%%%%%%%%%%%%%%%%%%%%%%%%%%%%%%%%%%%%%%%%%%%%%%%%%%%%%%%%%%%%%%%%%%%%%%%%%%%%%%%%%%%%%%%%%%%%%%%%%%%%%%%%%%%%%%%%%%
%%%%%%%%%%%%%%%%%%%%%%%%%%%%%%%%%%%%%%%%%%%%%%%%%%%%%%%%%%%%%%%%%%%%%%%%%%%%%%%%%%%%%%%%%%%%%%%%%%%%%%%%%%%%%%%%%%%%%%%%
%%%%%% INTRODUCTION %%%%%%
%%%%%%%%%%%%%%%%%%%%%%%%%%%%%%%%%%%%%%%%%%%%%%%%%%%%%%%%%%%%%%%%%%%%%%%%%%%%%%%%%%%%%%%%%%%%%%%%%%%%%%%%%%%%%%%%%%%%%%%%
%%%%%%%%%%%%%%%%%%%%%%%%%%%%%%%%%%%%%%%%%%%%%%%%%%%%%%%%%%%%%%%%%%%%%%%%%%%%%%%%%%%%%%%%%%%%%%%%%%%%%%%%%%%%%%%%%%%%%%%%%%%%%%%%%
%%%%%%%%%%%%%%%%%%%%%%%%%%%%%%%%%%%%%%%%%%%%%%%%%%%%%%%%%%%%%%%%%%%%%%%%%%%%%%%%%%%%%%%%%%%%%%%%%%%%%%%%%%%%%%%%%%%%%%%%%%%%%%%%%
\section{Introduction}\label{S1}
{\ Tremendous efforts inside the information/communication theory communities have been conducted in order to better understand performance limits of wireless communications in the low power regime. In this letter, we study the capacity of point-to-point fading channels with full channel state information at both the transmitter and the receiver (CSI-TR), as a performance limit at low SNR.
Indeed, many wireless systems should operate at low-SNR (equivalently at low-power), and hence the interest of deriving performance limits of communication at low-SNR~\cite{Verduspectral,spectraleff,widebandfading,multipleantenna,nakagamirez,lowsnrmimo,lowsnrjin,lowsnrcoherence}.
\par}
{\ For fading channels with perfect CSI at the receiver (CSI-R), a low SNR framework has been investigated in~\cite{multipleantenna,lowsnrmimo,Verduspectral}. A study of the effect of channel coherence on the capacity and energy efficiency of non-coherent fading channels with perfect CSI-TR at low SNR has been conducted in~\cite{lowsnrcoherence,Verduspectral}. In~\cite{widebandfading}, Borade and Zheng have shown that the capacity of the Rayleigh flat fading channel at low SNR essentially scales as $\text{SNR}\log(1/\text{SNR})$ nats/symbol. Motivated by these results, we propose to investigate the capacity of an independent identically distributed (i.i.d.) flat MRC Rician fading channel with perfect CSI-R, and possibly imperfect CSI at the transmitter (CSI-T). The ergodic capacity of this channel has been widely investigated in the literature in order to derive closed form expressions and/or accurate approximations,~\cite{mimorician2mrc}. But, the exact capacity expression involves a Lagrange multiplier that depends on the transmit power constraint (defined as SNR in this letter)~\cite{spectraleff}. It is a priori not clear how does the Lagrange multiplier scales with SNR. We need to perform more sophisticated computer simulations to numerically evaluate the capacity. Moreover, numerical techniques do not provide the whole picture about the capacity and require extensive parameterized simulations that are time-consuming and difficult to implement. Instead, one can focus on asymptotic analysis that can give a better understanding of the capacity and this relies on the continuity and the smoothness of the capacity in function of SNR. 
%But, the exact capacity involves a Lagrange multiplier that depends on the transmit power constraint (defined as SNR in this letter)~\cite{spectraleff}. It is a priori not clear how does the Lagrange multiplier scales with SNR.
Only recently, the low-SNR regime capacity of a Multiple-Input Multiple-Output (MIMO) Rician channel has been looked at in~\cite{lowsnrmimo,lowsnrjin}, but assuming no CSI-T or just mean CSI, respectively.
\par}
%%%%%%
%%%%%
{\ In this letter, we focus on a Single-Input Multiple-Output (SIMO) Rician fading channel where the receiver performs a MRC combining technique and provide the asymptotically low-SNR characterization of the capacity. As shown below, our results in this paper extend in a non-trivial manner the ones for the Single-Input Single-Output (SISO) Rician fading channel in~\cite{mrcricianrez} to a SIMO channel. We also propose an on-off scheme that does not require perfect CSI-T and show that one bit feedback is enough to achieve the asymptotic capacity. This justifies our "possibly imperfect CSI-T" assumption.
\par}
%%%%
%%%%%%%%%%%%%%%%%%%%%%%%%%%%%%%%%%%%%%%%%%%%%%%%%%%%%%%%%%%%%%%%%%%%%%%%%%%%%%%%%%%%%%%%%%%%%%%%%%%%%%%%%%%%%%%%%%%%%%%%
%%%%%%%%%%%%%%%%%%%%%%%%%%%%%%%%%%%%%%%%%%%%%%%%%%%%%%%%%%%%%%%%%%%%%%%%%%%%%%%%%%%%%%%%%%%%%%%%%%%%%%%%%%%%%%%%%%%%%%%%
%%%%%% SYSTEM MODEL %%%%%%
%%%%%%%%%%%%%%%%%%%%%%%%%%%%%%%%%%%%%%%%%%%%%%%%%%%%%%%%%%%%%%%%%%%%%%%%%%%%%%%%%%%%%%%%%%%%%%%%%%%%%%%%%%%%%%%%%%%%%%%%
%%%%%%%%%%%%%%%%%%%%%%%%%%%%%%%%%%%%%%%%%%%%%%%%%%%%%%%%%%%%%%%%%%%%%%%%%%%%%%%%%%%%%%%%%%%%%%%%%%%%%%%%%%%%%%%%%%%%%%%%%%%%%%%%%
%%%%%%%%%%%%%%%%%%%%%%%%%%%%%%%%%%%%%%%%%%%%%%%%%%%%%%%%%%%%%%%%%%%%%%%%%%%%%%%%%%%%%%%%%%%%%%%%%%%%%%%%%%%%%%%%%%%%%%%%%%%%%%%%%
\section{System Model}\label{S2}
%%%%%
{\ We consider an i.i.d. flat Rician fading channel described as
\begin{IEEEeqnarray}{rCl}\label{channel1}
\boldsymbol{y}=\boldsymbol{h}x+\boldsymbol{v},
\end{IEEEeqnarray}
where $x$ is a complex random variable (RV) that represents the channel input, $\boldsymbol{y}$ is an \textit{L}-dimensional complex random vector that represents the channel output, $\boldsymbol{v}$ is an \textit{L}-dimension vector of complex random variables that represents the additive white Gaussian noise (AWGN) with a zero mean and a variance $N_0$, that is $ \boldsymbol{v} \sim CN(0,N_0) $. In (\ref{channel1}), $\boldsymbol{h}$ is an \textit{L}-dimensional complex vector that represents the Rician fading channel gain and can be modeled by
\begin{IEEEeqnarray}{rCl}
\boldsymbol{h}=\sqrt{\frac{K}{1+K}} \boldsymbol{\overline{h}} + \sqrt{\frac{1}{1+K}} \boldsymbol{h}_{\omega},
\end{IEEEeqnarray}
where $K>0$ is the Rician factor, $L$ is the number of diversity branches, $ \boldsymbol{\overline{h}} \in C^{L} $ is a normalized constant vector representing the line of sight component, and $\boldsymbol{h}_{\omega}$ is an $L$-dimensional circularly symmetric complex Gaussian vector with mean $0_{L}$ and covariance matrix $I_{L}$ ($0_L$ denotes the zero vector and $I_L$ denotes the $L\times L$ identity matrix).
We assume perfect CSI-R and possibly imperfect CSI-T. The source is constrained to an average power $ E_{x|\boldsymbol{h}}\left[\vert x\vert^{2} \right] \leq P_{avg} $. We have normalized the noise ($N_0=1$), without loss of generality, so that $P_{avg}$ can be seen as the average transmit SNR. As the transmitter knows fully the CSI, we average over the distribution of x conditioned on $h$. We mainly focus on asymptotically low SNR regime. In the remaining of this letter, we define $f(x) \approx g(x) $ if and only if $ \lim_{x \to 0} \frac{f(x)}{g(x)} = 1$. We denote by $P(.)$ the instantaneous optimal power function and by $\text{Prob}(.)$ the probability of a continuous random variable.% Inequalities $\lesssim$ and $\gtrsim$ are defined in the same way.
\par} %%%%%%%%%%%%%%%%%%%%%%%%%%%%%%%%%%%%%%%%%%%%%%%%%%%%%%%%%%%%%%%%%%%%%%%%%%%%%%%%%%%%%%%%%%%%%%%%%%%%%%%%%%%%%%%%%%%%%%%%
%%%%%%%%%%%%%%%%%%%%%%%%%%%%%%%%%%%%%%%%%%%%%%%%%%%%%%%%%%%%%%%%%%%%%%%%%%%%%%%%%%%%%%%%%%%%%%%%%%%%%%%%%%%%%%%%%%%%%%%%
%%%%%% SECTION 1: Low-SNR Capacity With Perfect CSI-T %%%%%%
%%%%%%%%%%%%%%%%%%%%%%%%%%%%%%%%%%%%%%%%%%%%%%%%%%%%%%%%%%%%%%%%%%%%%%%%%%%%%%%%%%%%%%%%%%%%%%%%%%%%%%%%%%%%%%%%%%%%%%%%
%%%%%%%%%%%%%%%%%%%%%%%%%%%%%%%%%%%%%%%%%%%%%%%%%%%%%%%%%%%%%%%%%%%%%%%%%%%%%%%%%%%%%%%%%%%%%%%%%%%%%%%%%%%%%%%%%%%%%%%%%%%%%%%%%
%%%%%%%%%%%%%%%%%%%%%%%%%%%%%%%%%%%%%%%%%%%%%%%%%%%%%%%%%%%%%%%%%%%%%%%%%%%%%%%%%%%%%%%%%%%%%%%%%%%%%%%%%%%%%%%%%%%%%%%%%%%%%%%%%
%%%%%%%%%%%%%%%%%%%%%%%%%%%%%%%%%%%%%%%%%%%%%%%%%%%%%%%%%%%%%%%%%%%%%%%%%%%%%%%%%%%%%%%%%%%%%%%%%%%%%%%%%%%%%%%%%%%%%%%%%%%%%%%%%
\section{Low-SNR Capacity With Perfect CSI-TR}\label{S4}
%%%%%%%%%%%%%%%%%%%%%%%%%%%%%%%%%%%%%%%%%%%%%%%%%%%%%%%%%%%%%%%%%%%%%%%%%%%%%%%%%%%%%%%%%%%%%%%%%%%%%%%%%%%%%%%%%%%%%%%%
%%%%%%%%%%%%%%%%%%%%%%%%%%%%%%%%%%%%%%%%%%%%%%%%%%%%%%%%%%%%%%%%%%%%%%%%%%%%%%%%%%%%%%%%%%%%%%%%%%%%%%%%%%%%%%%%%%%%%%%%
%%%%%% SUBSECTION 1: CAPACITY RESULTS %%%%%%
%%%%%%%%%%%%%%%%%%%%%%%%%%%%%%%%%%%%%%%%%%%%%%%%%%%%%%%%%%%%%%%%%%%%%%%%%%%%%%%%%%%%%%%%%%%%%%%%%%%%%%%%%%%%%%%%%%%%%%%%
%%%%%%%%%%%%%%%%%%%%%%%%%%%%%%%%%%%%%%%%%%%%%%%%%%%%%%%%%%%%%%%%%%%%%%%%%%%%%%%%%%%%%%%%%%%%%%%%%%%%%%%%%%%%%%%%%%%%%%%%%%%%%%%%%
%%%%%%%%%%%%%%%%%%%%%%%%%%%%%%%%%%%%%%%%%%%%%%%%%%%%%%%%%%%%%%%%%%%%%%%%%%%%%%%%%%%%%%%%%%%%%%%%%%%%%%%%%%%%%%%%%%%%%%%%%%%%%%%%%
%%%%%%%%%%%%%%%%%%%%%%%%%%%%%%%%%%%%%%%%%%%%%%%%%%%%%%%%%%%%%%%%%%%%%%%%%%%%%%%%%%%%%%%%%%%%%%%%%%%%%%%%%%%%%%%%%%%%%%%%%%%%%%%%%
\subsection{A General Capacity Results}
{\ Having perfect CSI, the receiver performs a MRC and converts the channel into a SISO channel as
\begin{IEEEeqnarray}{rCl}\label{chanMRC}
z=\boldsymbol{h}^{H} \boldsymbol{y}= \boldsymbol{h}^{H} \boldsymbol{h} \boldsymbol{x} + \boldsymbol{h}^{H} \boldsymbol{v}.
\end{IEEEeqnarray}
The MRC combining diversity provides the highest output SNR and the instantaneous power of channel gain after performing MRC is then equal to $\gamma= \vert \boldsymbol{h}^{H} \boldsymbol{h} \vert=\Vert \boldsymbol{h}\Vert^{2}$. It can be shown that the probability density function (PDF) of $\gamma$, is a noncentral Chi-square distribution with $2L$ degree of freedom and noncentrality parameter $\frac{K}{K+L}$ which is given by~\cite{mobileradio}
\begin{IEEEeqnarray}{rCl}\label{pdf1}
f_{\gamma}(x)= \left(\frac{L+K}{L \Omega}\right)^{\frac{L+1}{2}} \left(\frac{x}{K}\right)^{\frac{L-1}{2}} e^{-\frac{(L+K)x}{L \Omega}-K } I_{L-1}\left( \sqrt{\frac{4K(K+L)x}{L \Omega}} \right) 
\end{IEEEeqnarray}
where 
$\Omega=\frac{E[\gamma]}{L}$ is the expected value of the channel gain per branch and $I_{L-1}(.)$ is the modified Bessel function of the first kind and order $(L-1)$. The instantaneous optimal power $P(\gamma)$ satisfying the average power constraint with equality $\displaystyle{P_{avg}=\text{SNR}= E_{\gamma}\left[ P(\gamma)\right]}$ is obtained by solving the Karush-Kuhn-Tucker (KKT) equation and is given by the water-filling policy~\cite{Tse}
\begin{IEEEeqnarray}{rCl}\label{policy}
P(\gamma)= \left[ \frac{1}{\lambda(\text{SNR})} - \frac{1}{\gamma} \right]^{+},
\end{IEEEeqnarray}
where $\lambda(\text{SNR})$ is the Lagrange multiplier. 
Let us define at this point the function $G(x)$ over $(0, \infty )$ as 
$\displaystyle{G(x)=E_{\gamma}\left[\left[ \frac{1}{x} - \frac{1}{\gamma} \right]^{+}\right]}$, so that
\begin{IEEEeqnarray}{rCl}
G(\lambda(\text{SNR}))&=&P_{avg}=\text{SNR}.\label{powerint}
\end{IEEEeqnarray}
The ergodic capacity is then obtained by averaging $\displaystyle{\log(1+P(\gamma) \gamma)}$ over (\ref{pdf1}) yielding
\begin{IEEEeqnarray}{rCl}
C(\lambda(\text{SNR}))
&=& \int_{\lambda}^{\infty} \! \log\left(\frac{x}{\lambda(\text{SNR})}\right) f_{\gamma}(x) \, \mathrm{d} x.\label{capaint}%)&=& \int_{0}^{\infty} \! \log\left(1+\left[ \frac{1}{\lambda(\text{SNR})} - \frac{1}{x} \right]^{+}x \right) f_{\gamma}(x) \, \mathrm{d} x,\nonumber\\
\end{IEEEeqnarray}
\theorem{For an i.i.d. flat Rician fading channel described by (\ref{channel1}), with perfect CSI at both the transmitter and the receiver, the low-SNR capacity is given by
\begin{IEEEeqnarray}{rCl}
C(\text{SNR})&\approx & 
\begin{cases} 
\frac{L \Omega (3-L)}{K+L} \hspace{1mm} \text{SNR}\hspace{1mm} W_{0}\left(\left(\frac{1}{\text{SNR}}\right)^{\frac{1}{3-L}}\right), & \mbox{if } L < 3, \\
\frac{3 \Omega}{K+3} \hspace{1mm} SNR \hspace{1mm} \log{\left(\frac{1}{\text{SNR}}\right)}, & \mbox{if } L = 3,\\ 
\frac{L \Omega(3-L)}{K+L} \hspace{1mm} \text{SNR}\hspace{1mm} W_{-1}\left(-\left(\frac{1}{\text{SNR}}\right)^{\frac{1}{3-L}}\right), & \mbox{if } L > 3,
\end{cases}\label{theoremcapacity}\\
&\approx & \frac{L \Omega}{K+L} \hspace{1mm} \text{SNR}\hspace{1mm} \log{\left(\frac{1}{\text{SNR}}\right)},\label{theoremcapacitytotal}
\end{IEEEeqnarray}
where $W_{0}(.)$ and $W_{-1}(.)$ are the principal branch and the lower branch of the Lambert function, respectively.}
\proof{Using Lemma 1 in~\cite{nakagamirez}, the function $G(.)$ is continuous, positive definite, and strictly monotonically increasing. Thus, we can show~\cite{nakagamirez} that $\underset{\text{SNR}\rightarrow 0}{\lim} \lambda(\text{SNR}) = +\infty$. Using this fact, we use the series expansion of $\displaystyle{I_{L-1}(x)}$ at infinity~\cite{series}:
$\displaystyle{I_{L-1}(x)= \sum_{p=0}^{\infty} \frac{(x/2)^{L-1+2p}}{p! (L+p-1)!}}$,
and substituting it in the expression of the PDF given in (\ref{pdf1}) yields
\begin{IEEEeqnarray}{rCl}\label{pdfser}
f_{\gamma}(x)&\approx & e^{-K} \sum_{p=0}^{\infty} \frac{K^p (\frac{K+L}{L \Omega})^{p+L}}{p! (L+p-1)!} x^{L+p-1}e^{-\frac{(L+K)x}{L \Omega}}.
\end{IEEEeqnarray}
Substituting (\ref{pdfser}) in (\ref{capaint}) and (\ref{powerint}), we can then write
\begin{IEEEeqnarray}{rCl}
C(\lambda) &\approx & e^{-K} \sum_{p=0}^{\infty} \frac{K^p (\frac{K+L}{L \Omega})^{p+L}}{p! (L+p-1)!} \int_{\lambda}^{\infty} \log\left(\frac{x}{\lambda}\right) x^{L+p-1}e^{-\frac{(L+K)x}{L \Omega}} \, \mathrm{d} x,\nonumber\\
&\approx & e^{-K} e^{-\frac{(L+K)\lambda}{L \Omega}} \lambda^L \times \left[\frac{(\frac{K+L}{L \Omega})^{L-2}}{(L-1)! \lambda^2} + \frac{(2L-3)(\frac{K+L}{L \Omega})^{L-3}}{(L-1)!\lambda^3}+ o\left(\frac{1}{\lambda^4}\right)\right],\nonumber\\
&\approx & e^{-K} \frac{(\frac{K+L}{L \Omega})^{L-2}}{(L-1)!} e^{-\frac{(L+K)\lambda}{L \Omega}} \lambda^{L-2},\label{capacityfunlambdaexpansion}
\end{IEEEeqnarray}
\begin{IEEEeqnarray}{rCl}
\text{SNR}&\approx & 
e^{-K} \sum_{p=0}^{\infty} \frac{K^p (\frac{K+L}{L \Omega})^{p+L}}{p! (L+p-1)!} \int_{\lambda}^{\infty} \left[\frac{1}{\lambda}-\frac{1}{x}\right] x^{L+p-1}e^{-\frac{(L+K)x}{L \Omega}} \, \mathrm{d} x, \nonumber\\
&\approx & e^{-K} e^{-\frac{(L+K)\lambda}{L \Omega}} \lambda^L \left[\frac{(\frac{K+L}{L \Omega})^{L-2}}{(L-1)! \lambda^3}+ o\left(\frac{1}{\lambda^4}\right)\right],\nonumber\\
&\approx & \frac{e^{-K}(\frac{K+L}{L \Omega})^{L-2}}{(L-1)!} e^{-\frac{(L+K)}{L \Omega}\lambda} \lambda^{L-3}.\label{SNRfunlambdaexpansion}
\end{IEEEeqnarray}
Eq. (\ref{SNRfunlambdaexpansion}) is of the form $y=xe^x$ and its solution depends on the value of $L$ as follows:
\begin{itemize}
\item If $L=3$, then a solution of Eq. (\ref{SNRfunlambdaexpansion}) can be found by solving the equation: 
$\displaystyle{\text{SNR}\approx \frac{e^{-K} (\frac{K+3}{L \Omega})}{2} e^{-\frac{(K+3)}{L \Omega}\lambda}}$.
We can then write
\begin{IEEEeqnarray}{rCl}
\lambda(\text{SNR}) &\approx & \frac{L \Omega}{(K+3)} \log\left(\frac{(\frac{K+3}{L \Omega})e^{-K}}{2 SNR}\right).\label{capacityleq3withconst}
\end{IEEEeqnarray}
Note that Eq. (\ref{capacityleq3withconst}) is valid only for $\displaystyle{ \text{SNR}\leq \frac{L+K}{2 L \Omega }e^{-K}}$.
\item If $L<3$, then a solution of Eq. (\ref{SNRfunlambdaexpansion}) can be expressed using the principal branch of the Lambert function and is given by
\begin{IEEEeqnarray}{rCl}
\lambda(\text{SNR}) &\approx & \frac{L \Omega(3-L)}{K+L} W_{0}\left( \alpha \hspace{1mm}\left(\frac{1}{ \text{SNR}} \right)^{\frac{1}{3-L}}\right)\label{capacitylinf3withconst},
\end{IEEEeqnarray}
with $\displaystyle{\alpha=\frac{1}{3-L} \left(\frac{e^{-K} {\left(\frac{K+L}{L\Omega}\right)}}{(L-1)!} \right)^{\frac{1}{3-L}}}$.
\item If $L>3$, Eq. (\ref{SNRfunlambdaexpansion}) can be solved using the lower branch of the Lambert function yields
\begin{IEEEeqnarray}{rCl}
\lambda(\text{SNR}) &\approx & \frac{L \Omega(3-L)}{K+L} W_{-1}\left( \alpha \hspace{1mm} \left(\frac{1}{\text{SNR}} \right)^{\frac{1}{3-L}}\right).\label{capacitylsup3withconst}
\end{IEEEeqnarray}
Note that the argument in (\ref{capacitylsup3withconst}) should be greater than $(-\frac{1}{e})$. That is, we can get valid values of $\lambda(\text{SNR})$ that satisfies (\ref{powerint}) only for values of $\text{SNR}$ less than $\left(\frac{-\alpha}{e}\right)^{L-3}$~\cite{tall}.
\end{itemize}
On the other hand, using Eqs. (\ref{capacityfunlambdaexpansion}) and (\ref{SNRfunlambdaexpansion}), we can write
\begin{IEEEeqnarray}{rCl}
C(\text{SNR}) &\approx & \text{SNR}\hspace{1mm} \lambda(\text{SNR}).
\end{IEEEeqnarray}
To complete the proof of (\ref{theoremcapacity}), we use l'H$\hat{o}$pital rule to show that for any $\beta >0$, $x>0$ and $y<0$ we have~\cite{nakagamirez}\\
\begin{center}
$\underset{x \rightarrow\infty}{\lim} \frac{W_{0}(\beta x)}{W_{0}(x)} = 1$ and
$\underset{y \rightarrow 0^{-}}{\lim} \frac{W_{-1}(\beta y)}{W_{-1}(y)} = 1$.
\end{center}
Thus, Eqs. (\ref{capacitylinf3withconst}) and (\ref{capacitylsup3withconst}) simplify to
\begin{IEEEeqnarray}{rCl}
\lambda(\text{SNR}) &\approx & \frac{L \Omega(3-L)}{K+L} W_{0}\left(\left(\frac{1}{\text{SNR}} \right)^{\frac{1}{3-L}}\right)\label{capacitylinf3},
\end{IEEEeqnarray}
and
\begin{IEEEeqnarray}{rCl}
\lambda(\text{SNR}) &\approx & \frac{L \Omega(3-L)}{K+L} W_{-1}\left( - \left(\frac{1}{\text{SNR}} \right)^{\frac{1}{3-L}}\right),\label{capacitylsup3}
\end{IEEEeqnarray}
respectively. As shown in~\cite{nakagamirez}, we can approximate the Lambert function by a familiar function $(\log(.))$ since $\displaystyle{\underset{x \rightarrow\infty}{\lim} \frac{W_{0}(x)}{\log(x)}=1}$ and $\displaystyle{\underset{y \rightarrow0^{-}}{\lim} \frac{W_{-1}(y)}{\log(-y)}=1}$, for $x>0$ and $y<0$, and using the fact that $\lambda(\text{SNR})\rightarrow \infty$ as $\text{SNR}\rightarrow 0$ we can get immediately the simpler result given in (\ref{theoremcapacitytotal}).
}\par}
%%%%%%%%%%%%%%%%%%%%%%%%%%%%%%%%%%%%%%%%%%%%%%%%%%%%%%%%%%%%%%%%%%%%%%%%%%%%%%%
%%%%%%%%%%%%%%%%%%%%%%%%%%%%%%%%%%%%%%%%%%%%%%%%%%%%%%%%%%%%%%%%%%%%%%%%%%%%%%%%%
%%%%%%%%%%%%%%%%%%%%%%%%%%%%%%%%%%%%%%%%%%%%%%%%%%%%%%%%%%%%%%%%%%%%%%%%%%%%%%%%%%%
\subsection{Special Cases And Remarks}
{\ We use the results derived in the previous subsection to treat some interesting and non-trivial special cases and also to compute the energy efficiency at low SNR:
\begin{itemize}
\item[1)] When $K$ tends to infinity, the flat Rician fading channel becomes equivalent to an $L$-branch AWGN channel and the PDF $\displaystyle{ f_{\Vert \boldsymbol{h}\Vert^{2}}(x) \rightarrow \delta(x-L \Omega) }$, where $\delta(.)$ is the Dirac-delta function. Hence, the capacity simplifies to
\begin{IEEEeqnarray}{rCl}
C(\text{SNR})&=& \int_{0}^{\infty} \log\left(1+\left[\frac{1}{\lambda}-\frac{1}{t}\right]^{+}t\right) \delta(t-L \Omega) \, \mathrm{d} t,\\
&=& \log\left(1+\left[\frac{1}{\lambda}-\frac{1}{L \Omega}\right]^{+} L \Omega\right),\\
&\approx& \left[\frac{1}{\lambda}-\frac{1}{L \Omega}\right]^{+} L\hspace{1mm} \Omega = L\hspace{1mm} \Omega\hspace{1mm}\text{SNR}.\label{Casinf}
\end{IEEEeqnarray}
\item[2)] By letting $L$ tends toward infinity in Eq. (\ref{theoremcapacitytotal}), it can be seen that the capacity scales linearly with $L$ as $ L \hspace{1mm}\Omega\hspace{1mm} \text{SNR}$ at asymptotically low SNR. This result can be retrieved using a similar reasoning than the one above.
\item[3)] The energy efficiency of a flat Rician fading channel at low SNR can be characterized, given Eq. (\ref{theoremcapacitytotal}), by
\begin{IEEEeqnarray}{rCl}
\frac{E_{n}}{\sigma^{2}_{v}} = \frac{\text{SNR}}{C(\text{SNR})} \approx \frac{\frac{K+L}{L \Omega}}{\log\left(\frac{1}{\text{SNR}}\right)}.\label{energy}
\end{IEEEeqnarray}
where $E_{n}$ is the transmitted energy in Joules per information nats and $\sigma^{2}_{v}$ is the noise variance. Equation (\ref{energy}) states that communicating one nat of information reliably requires a very low energy when in addition to CSI-R, CSI-T is also available. Moreover, Eq. (\ref{energy}) shows the dependence of the energy efficiency on the number of diversity branches $L$ and the Rician factor $K$. Note that when only CSIR is available, the energy efficiency is equal to:
\begin{IEEEeqnarray}{rCl}\label{energy0}
\frac{E_{n}}{\sigma^2_v} &\approx \frac{\text{SNR}}{\text{E}[\log(1+\text{SNR}\hspace{1mm}\gamma )]} \approx \frac{1}{ \text{E}[\gamma]}=\frac{1}{L\Omega},
\end{IEEEeqnarray}
a constant regardless of the SNR value.
\end{itemize}
\par}
%%%%%%%%%%%%%%%%%%%%%%%%%%%%%%%%%%%%%%%%%%%%%%%%%%%%%%%%%%%%%%%%%%%%%%%%%%%%%%%%%%%%%%%%%%%%%%%%%%%%%%%%%%%%%%%%%%%%%%%%
%%%%%%%%%%%%%%%%%%%%%%%%%%%%%%%%%%%%%%%%%%%%%%%%%%%%%%%%%%%%%%%%%%%%%%%%%%%%%%%%%%%%%%%%%%%%%%%%%%%%%%%%%%%%%%%%%%%%%%%%
%%%%%% SUBSECTION 2: ON-OFF POWER CONTROL IS ASYMPTOTICALLY OPTIMAL %%%%%%
%%%%%%%%%%%%%%%%%%%%%%%%%%%%%%%%%%%%%%%%%%%%%%%%%%%%%%%%%%%%%%%%%%%%%%%%%%%%%%%%%%%%%%%%%%%%%%%%%%%%%%%%%%%%%%%%%%%%%%%%
%%%%%%%%%%%%%%%%%%%%%%%%%%%%%%%%%%%%%%%%%%%%%%%%%%%%%%%%%%%%%%%%%%%%%%%%%%%%%%%%%%%%%%%%%%%%%%%%%%%%%%%%%%%%%%%%%%%%%%%%%%%%%%%%%
%%%%%%%%%%%%%%%%%%%%%%%%%%%%%%%%%%%%%%%%%%%%%%%%%%%%%%%%%%%%%%%%%%%%%%%%%%%%%%%%%%%%%%%%%%%%%%%%%%%%%%%%%%%%%%%%%%%%%%%%%%%%%%%%%
\subsection{On-Off Power Control Is Asymptotically Optimal}
{\ In this subsection, we use the insight gained from our analytical result (Theorem 1) in order to design a practical scheme that is asymptotically
capacity-achieving. Let us consider an On-Off power control scheme that transmits whenever $\gamma \geq \lambda(\text{SNR})$ and remains silent otherwise. Thus, $P(\gamma)$ is given by
\begin{IEEEeqnarray}{rCl}\label{poweronoff}
P(\gamma) = 
\begin{cases} \frac{\text{SNR}}{\text{Prob}(\gamma \geq \lambda(\text{SNR}))}, & \mbox{if } \gamma \geq \lambda(\text{SNR}), \\ 0, & \mbox{otherwise,} 
\end{cases}
\end{IEEEeqnarray}
where $\lambda(\text{SNR})$ satisfies (\ref{SNRfunlambdaexpansion}). The rate achievable by this scheme is given by
\begin{IEEEeqnarray}{rCl}
R(\text{SNR})&=&E_{\gamma}\left[\log(1+P(\gamma)\gamma)\right],\\
&=&\int_{\lambda(\text{SNR})}^{\infty} \log\left(1+ t \frac{\text{SNR}}{\text{Prob}(\gamma \geq \lambda(\text{SNR}))}\right) f_{\gamma}(t) \, \mathrm{d} t,\nonumber\\
&\geq & \int_{\lambda(\text{SNR})}^{\infty} \log\left(1+ \lambda(\text{SNR}) \frac{\text{SNR}}{\text{Prob}(\gamma \geq \lambda(\text{SNR}))}\right)f_{\gamma}(t) \, \mathrm{d} t,\nonumber\\
&=& \log\left(1+ \frac{\lambda(\text{SNR}) \text{SNR}}{\text{Prob}(\gamma \geq \lambda(\text{SNR}))}\right) \text{Prob}(\gamma \geq \lambda(\text{SNR}))\label{rateonoff}.
\end{IEEEeqnarray}
On the other hand, we have:
\begin{IEEEeqnarray}{rCl}
\text{Prob}(\gamma \geq \lambda(\text{SNR})) &=& \text{Prob}\left(\left\Vert\sqrt{\frac{K}{1+K}} \overline{h} + \sqrt{\frac{1}{1+K}} \boldsymbol{h}_{\omega}\right\Vert^{2} \geq \lambda(\text{SNR})\right),\nonumber\\
&\approx & \text{Prob}\left(\left\Vert\boldsymbol{h}_{\omega}\right\Vert^{2} \geq (1+K)\hspace{1mm} \lambda(\text{SNR})\right),\label{Raylei}\\
&= & \frac{\Gamma\left(L,\frac{(K+1)\lambda(\text{SNR})}{L \Omega}\right)}{(L-1)!},\label{gamma}\\
&\approx & \frac{\left(\frac{(K+1)\lambda(\text{SNR})) }{L \Omega}\right)^{L-1} e^{-\frac{(K+1)\lambda(\text{SNR})}{L \Omega}}}{(L-1)!},\label{gammaapprox}
\end{IEEEeqnarray}
where $\Gamma(.,.)$ is the incomplete gamma function. Eq. (\ref{Raylei}) is due to the fact that the line of sight component $\sqrt{\frac{K}{K+1}}\overline{h}$ is neglected as $\text{SNR}\rightarrow 0$ since $\lambda \rightarrow \infty$. 
Eq. (\ref{gammaapprox}) is derived from (\ref{gamma}) using the fact that $\underset{x \rightarrow\infty}{\lim} \Gamma(s,x) \approx x^{s-1} e^{-x}$. Thus, 
\begin{IEEEeqnarray}{rCl}
\frac{\lambda(\text{SNR})\text{SNR}}{\text{Prob}(\gamma \geq \lambda(\text{SNR}))}
&\approx & e^{-K} \frac{L \Omega }{K+1} \left(\frac{K+L}{K+1}\right)^{L-2} \frac{e^{-\frac{(L-1)\lambda}{L \Omega}}}{ \lambda },\label{lim1}\\
&\underset{\lambda \rightarrow \infty}{\rightarrow}& 0.\label{lim0}
\end{IEEEeqnarray}
Eq. (\ref{lim0}) applies from Eq. (\ref{lim1}) since $L\geq 1$. Then, using (\ref{rateonoff}) and the asymptotic approximation $\log(x+1) \approx x$, we can deduce that the achievable rate of the on-off scheme is equal to $\lambda(\text{SNR})\hspace{1mm}\text{SNR}$. Note that given Eq.(\ref{SNRfunlambdaexpansion}) and Eq.(\ref{gammaapprox}), $P(\gamma)$ goes to zero as $\text{SNR}\rightarrow 0$, in agreement with Eq.(\ref{policy}) which also goes to zero as $\lambda$ goes to infinity. Note also that only 1-bit feedback of CSI-T in each fading realization is enough to achieve the asymptotic capacity. This bit contains the result of the outcome of the comparison between $\gamma$ and $\lambda(\text{SNR})$.
\par}
%%%%%%%%%%%%%%%%%%%%%%%%%%%%%%%%%%%%%%%%%%%%%%%%%%%%%%%%%%%%%%%%%%%%%%%%%%%%%%%%%%%%%%%%%%%%%%%%%%%%%%%%%%%%%%%%%%%%%%%%
%%%%%%%%%%%%%%%%%%%%%%%%%%%%%%%%%%%%%%%%%%%%%%%%%%%%%%%%%%%%%%%%%%%%%%%%%%%%%%%%%%%%%%%%%%%%%%%%%%%%%%%%%%%%%%%%%%%%%%%%
%%%%%% NUMERICAL RESULTS %%%%%%
%%%%%%%%%%%%%%%%%%%%%%%%%%%%%%%%%%%%%%%%%%%%%%%%%%%%%%%%%%%%%%%%%%%%%%%%%%%%%%%%%%%%%%%%%%%%%%%%%%%%%%%%%%%%%%%%%%%%%%%%
%%%%%%%%%%%%%%%%%%%%%%%%%%%%%%%%%%%%%%%%%%%%%%%%%%%%%%%%%%%%%%%%%%%%%%%%%%%%%%%%%%%%%%%%%%%%%%%%%%%%%%%%%%%%%%%%%%%%%%%%%%%%%%%%%
%%%%%%%%%%%%%%%%%%%%%%%%%%%%%%%%%%%%%%%%%%%%%%%%%%%%%%%%%%%%%%%%%%%%%%%%%%%%%%%%%%%%%%%%%%%%%%%%%%%%%%%%%%%%%%%%%%%%%%%%%%%%%%%%%
\section{Numerical Results}\label{S6}
%%%%
{\ In this section, some selected numerical results are provided to show the accuracy of our characterization. We have plotted in Figs. \ref{capacityL3K1} and \ref{capacityL2K2}, the ergodic capacity of an i.i.d. MRC Rician fading channel in nats per channel use (npcu) with perfect CSI-TR for $(L, K)$ equal to $(3,1)$ and $(2,2)$, respectively. The non-asymptotic curves have been obtained using standard optimization methods, whereas the asymptotic ones represent the low SNR characterizations given by (\ref{theoremcapacity}) and (\ref{theoremcapacitytotal}) in Theorem 1. In both figures, we can see that the curves depicting the characterizations in Theorem 1 follow the same shape as the curve obtained by simulations. In Fig. \ref{capacityL3K1}, we can see that the shape of the asymptotic results change for $\text{SNR}\geq -4$ $dB$ and this is because the slope of $\text{SNR}\log(1/\text{SNR})$ changes at $\text{SNR}=e^{-1}$ and becomes negative, independently of the value of $L$ and $K$. At low SNR regime which is the focus of this letter, and more specifically for $\text{SNR}\leq -4$ $dB$, our characterization is very accurate as shown in Fig. \ref{capacityL3K1}. In Fig. \ref{capacityL3K1}, the asymptotic characterizations (\ref{theoremcapacity}) and (\ref{theoremcapacitytotal}) are the same, since $L=3$. Note that our asymptotic characterizations are accurate for small values of $K$. Intuitively, as $K \rightarrow \infty$, the channel tends to be an AWGN channel and the capacity scales only proportionally with SNR, a known result that still can be retrieved by our framework, as discussed in B.1). In Figs. \ref{capacityL3K1} and \ref{capacityL2K2}, we have also plotted the asymptotic capacity as $L$ or $K$ $\rightarrow \infty$ given by (\ref{Casinf}). Furthermore, the on-off achievable rate is also depicted in Figs. \ref{capacityL3K1} and \ref{capacityL2K2}, where it can be seen that it is in fact the closest curve to the capacity suggesting that this suboptimal scheme is worth implementing in the low-SNR regime. %and we can see that our asymptotic capacity goes closer to the capacity for $K \rightarrow \infty$ or $L \rightarrow \infty$ as $L$ grows or $K$ grows. Figure \ref{energyL3K1} depicts the energy efficiency obtained with CSI-TR and the energy efficiency with CSI-R only, versus SNR for $L=3$ and $K=1$. The two curves with dashed lines in Fig. \ref{energyL3K1} have been obtained using the Eqs. (\ref{energy}) and (\ref{energy0}), whereas the other two curves have been obtained using exact values of the ergodic capacity. Figure \ref{energyL3K1} highlights the gain in terms of energy efficiency obtained through CSI-TR over CSI-R only, especially at low-SNR ($\text{SNR}\leq -20dB$).\par}
%%%%%%%%%%%%%%%%%%%%%%%%%%%%%%%%%%%%%%%%%%%%%%%%%%%%%%%%%%%%%%%%%%%%%%%%%%%%%%%%%%%%%%%%%%%%%%%%%%%%%%%%%%%%%%%%%%%%%%%%
%%%%%%%%%%%%%%%%%%%%%%%%%%%%%%%%%%%%%%%%%%%%%%%%%%%%%%%%%%%%%%%%%%%%%%%%%%%%%%%%%%%%%%%%%%%%%%%%%%%%%%%%%%%%%%%%%%%%%%%%
%%%%%% CONCLUSION %%%%%%
%%%%%%%%%%%%%%%%%%%%%%%%%%%%%%%%%%%%%%%%%%%%%%%%%%%%%%%%%%%%%%%%%%%%%%%%%%%%%%%%%%%%%%%%%%%%%%%%%%%%%%%%%%%%%%%%%%%%%%%%
%%%%%%%%%%%%%%%%%%%%%%%%%%%%%%%%%%%%%%%%%%%%%%%%%%%%%%%%%%%%%%%%%%%%%%%%%%%%%%%%%%%%%%%%%%%%%%%%%%%%%%%%%%%%%%%%%%%%%%%%%%%%%%%%%
%%%%%%%%%%%%%%%%%%%%%%%%%%%%%%%%%%%%%%%%%%%%%%%%%%%%%%%%%%%%%%%%%%%%%%%%%%%%%%%%%%%%%%%%%%%%%%%%%%%%%%%%%%%%%%%%%%%%%%%%%%%%%%%%%
\section{Conclusion}\label{S7}
{\ We have analyzed the capacity of an i.i.d. MRC Rician fading channel at low SNR for perfect CSI-TR and we have shown that it scales proportionally as $\text{SNR}\log(\frac{1}{\text{SNR}})$. An on-off power control that exploits the available CSI at the transmitter has been shown to be asymptotically optimal.Furthermore, the energy efficiency at low-SNR regime has been also characterized in terms of \text{SNR}, the number of diversity branches $L$ and the Rician factor $K$. Finally, numerical results have been provided to show the accuracy of our characterization. \par}
%%%%%%%%%%%%%%%%%%%%%%%%%%%%%%%%%%%%%%%%%%%%%%%%%%%%%%%%%%%%%%%%%%%%%%%%%%%%%%%%%%%%%%%%%%%%%%%%%%%%%%%%%%%%%%%%%%%%%%%%%
%%%%%%%%%%%%%%%%%%%%%%%%%%%%%%%%%%%%%%%%%%%%%%%%%%%%%%%%%%%%%%%%%%%%%%%%%%%%%%%%%%%%%%%%%%%%%%%%%%%%%%%%%%%%%%%%%%%%%%%%%
%%%%%%% BIBLIOGRAPHY %%%%%%
%%%%%%%%%%%%%%%%%%%%%%%%%%%%%%%%%%%%%%%%%%%%%%%%%%%%%%%%%%%%%%%%%%%%%%%%%%%%%%%%%%%%%%%%%%%%%%%%%%%%%%%%%%%%%%%%%%%%%%%%%
%%%%%%%%%%%%%%%%%%%%%%%%%%%%%%%%%%%%%%%%%%%%%%%%%%%%%%%%%%%%%%%%%%%%%%%%%%%%%%%%%%%%%%%%%%%%%%%%%%%%%%%%%%%%%%%%%%%%%%%%%%%%%%%%%%
%%%%%%%%%%%%%%%%%%%%%%%%%%%%%%%%%%%%%%%%%%%%%%%%%%%%%%%%%%%%%%%%%%%%%%%%%%%%%%%%%%%%%%%%%%%%%%%%%%%%%%%%%%%%%%%%%%%%%%%%%%%%%%%%%%
\bibliographystyle{IEEEtran} % style de la bibliographie
\bibliography{BenkhelifaRezkiAlouini_MRCRician_references}%Biblio/definition,
%%%%%%%%%%%%%%%%%%%%%%%%%%%%%%%%%%%%%%%%%%%%%%%%%%%%%%%%%%%%%%%%%%%%%%%%%%%%%%%%%%%%%%%%%%%%%%%%%%%%%%%%%%%%%%%%%%%%%%%%
%%%%%%%%%%%%%%%%%%%%%%%%%%%%%%%%%%%%%%%%%%%%%%%%%%%%%%%%%%%%%%%%%%%%%%%%%%%%%%%%%%%%%%%%%%%%%%%%%%%%%%%%%%%%%%%%%%%%%%%%
%%%%%% FIGURES %%%%%%
%%%%%%%%%%%%%%%%%%%%%%%%%%%%%%%%%%%%%%%%%%%%%%%%%%%%%%%%%%%%%%%%%%%%%%%%%%%%%%%%%%%%%%%%%%%%%%%%%%%%%%%%%%%%%%%%%%%%%%%%
%%%%%%%%%%%%%%%%%%%%%%%%%%%%%%%%%%%%%%%%%%%%%%%%%%%%%%%%%%%%%%%%%%%%%%%%%%%%%%%%%%%%%%%%%%%%%%%%%%%%%%%%%%%%%%%%%%%%%%%%%%%%%%%%%
%%%%%%%%%%%%%%%%%%%%%%%%%%%%%%%%%%%%%%%%%%%%%%%%%%%%%%%%%%%%%%%%%%%%%%%%%%%%%%%%%%%%%%%%%%%%%%%%%%%%%%%%%%%%%%%%%%%%%%%%%%%%%%%%%
%%%%%%% FIGURE 1 L=3, K=1
\begin{figure}[t]
\begin{center}
\includegraphics[scale=0.27]
{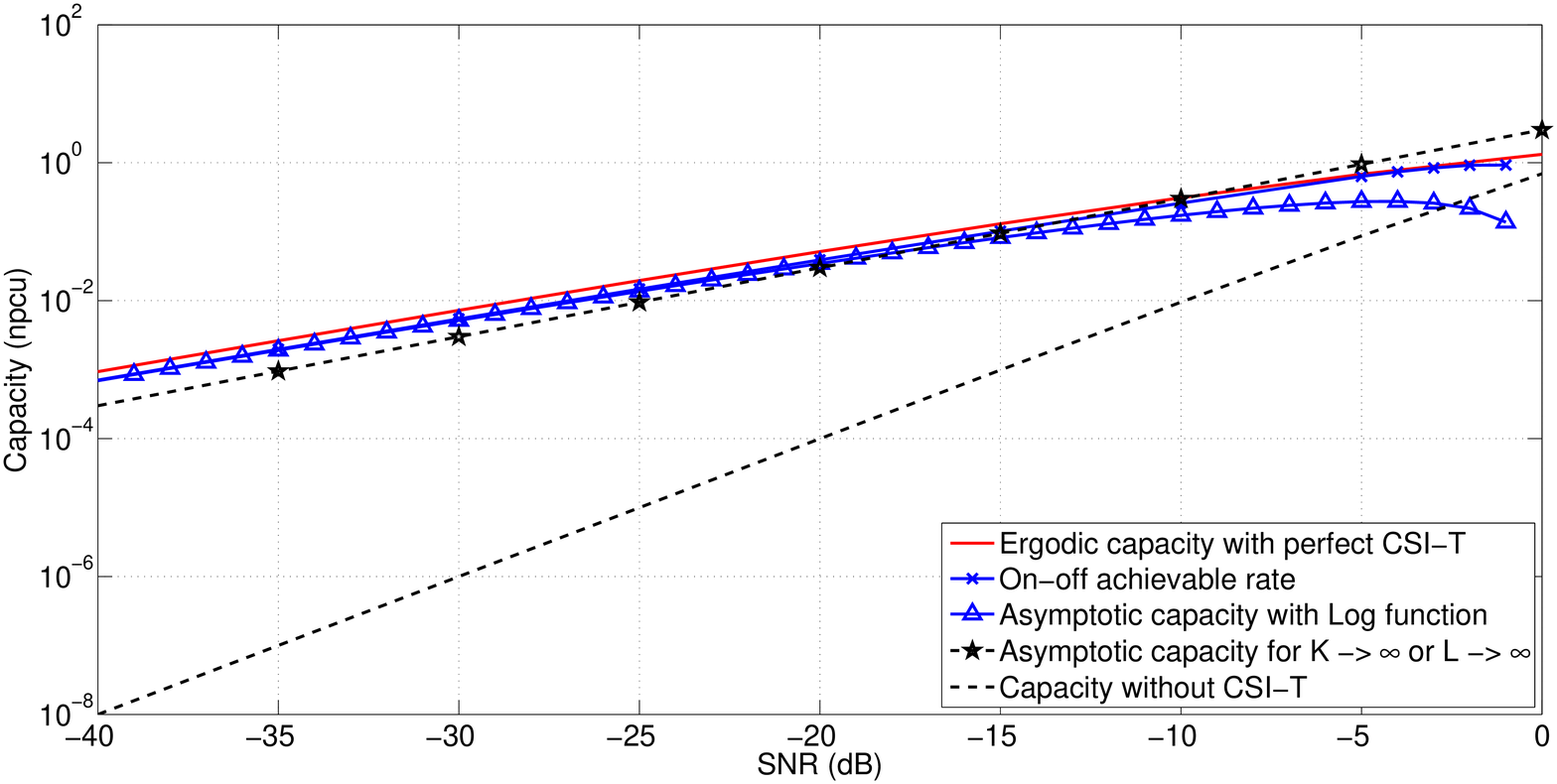}
\caption{Low-SNR capacity in nats per channel use (npcu) versus SNR for $L=3$ and $K=1$.}
\label{capacityL3K1}% ,height=4.1cm
%\end{center}
%\end{figure}
%%%%%%%% FIGURE 1 L=2, K=2
%\begin{figure}[t]
%\begin{center}
\includegraphics[scale=0.27]
{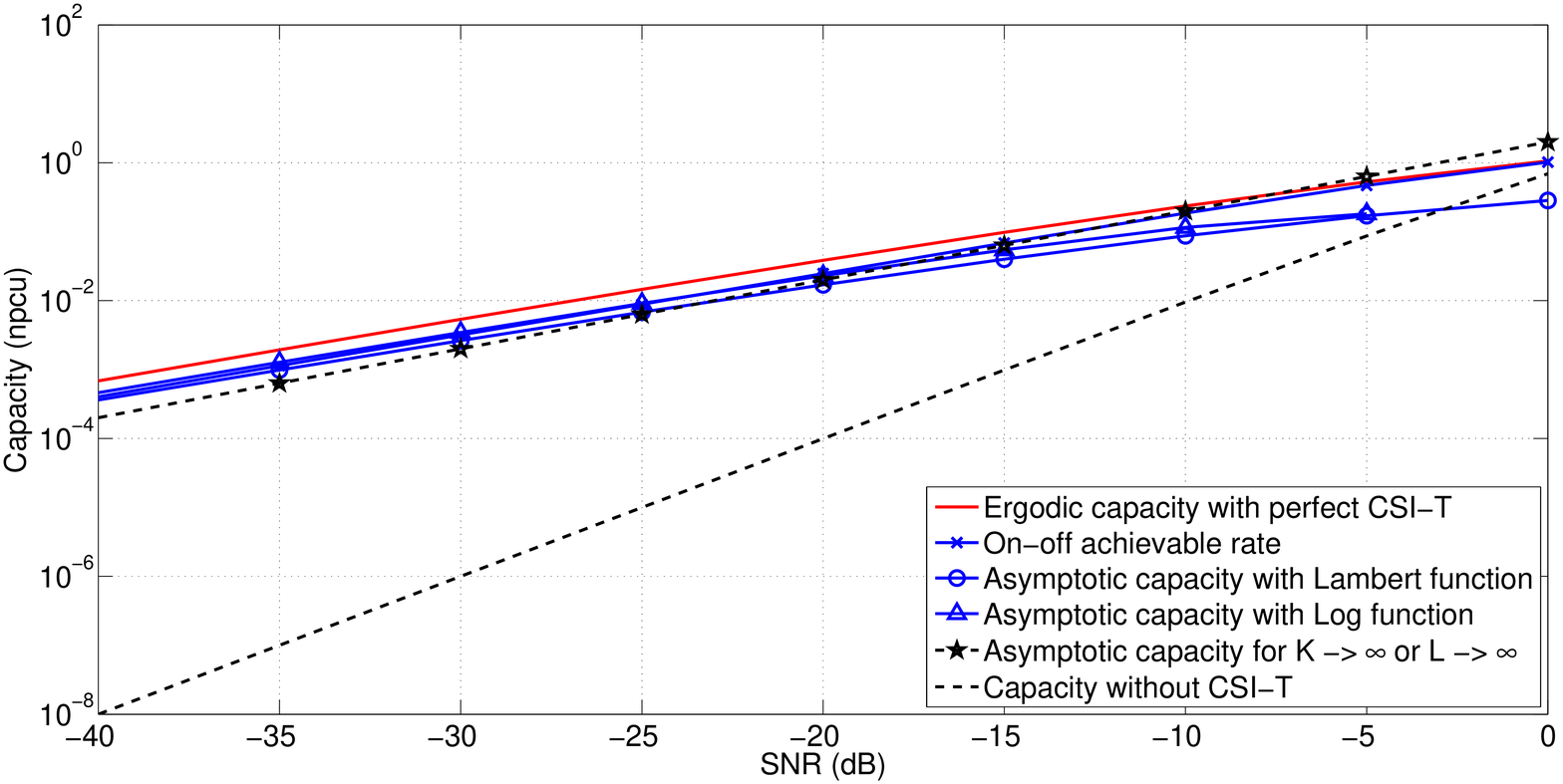}
\caption{Low-SNR capacity in nats per channel use (npcu) versus SNR for $L=2$ and $K=2$.}
\label{capacityL2K2}
%\end{center}
%\end{figure}
%%%%%%%%%%%% figure energy efficiency L=3,K=1
%\begin{figure}[t]
%\begin{center}
\includegraphics[scale=0.27]
{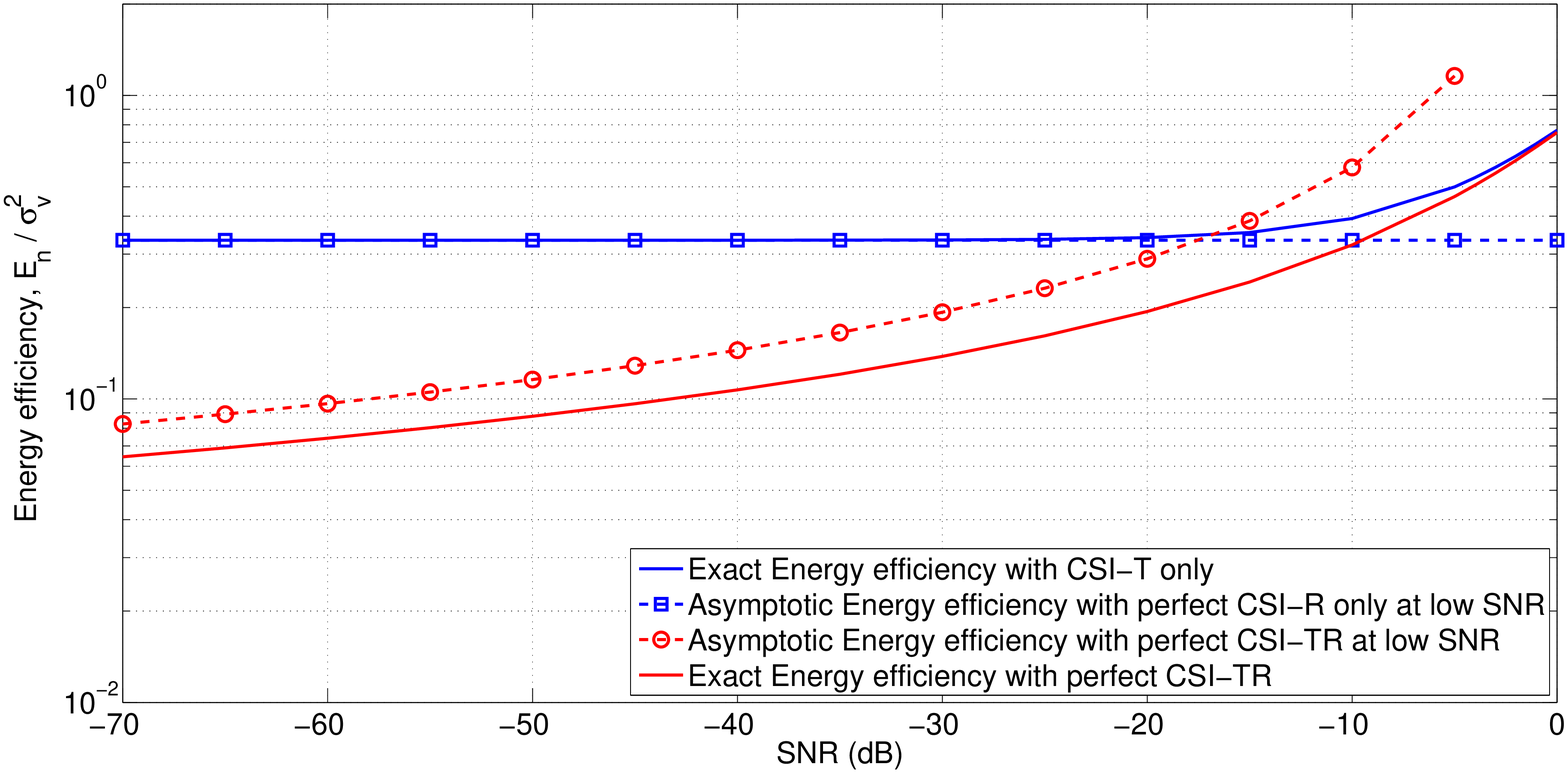}
\caption{Energy efficiency versus SNR for $L=3$ and $K=1$.}
\label{energyL3K1}
\end{center}
\end{figure}
%%%%%%%%%%%%%%%%%%%%%%%%%%%%%%%%%%%%%%%%%%%%%%%%%%%%%%%%%%%%%%%%%%%%%%%%%%%%%%%%%%%%%%%%%%%%%%%%%%%%%%%%%%%%%%%%%%%%%%%%
%%%%%%%%%%%%%%%%%%%%%%%%%%%%%%%%%%%%%%%%%%%%%%%%%%%%%%%%%%%%%%%%%%%%%%%%%%%%%%%%%%%%%%%%%%%%%%%%%%%%%%%%%%%%%%%%%%%%%%%%
%%%%%% END %%%%%%
%%%%%%%%%%%%%%%%%%%%%%%%%%%%%%%%%%%%%%%%%%%%%%%%%%%%%%%%%%%%%%%%%%%%%%%%%%%%%%%%%%%%%%%%%%%%%%%%%%%%%%%%%%%%%%%%%%%%%%%%
%%%%%%%%%%%%%%%%%%%%%%%%%%%%%%%%%%%%%%%%%%%%%%%%%%%%%%%%%%%%%%%%%%%%%%%%%%%%%%%%%%%%%%%%%%%%%%%%%%%%%%%%%%%%%%%%%%%%%%%%%%%%%%%%%
%%%%%%%%%%%%%%%%%%%%%%%%%%%%%%%%%%%%%%%%%%%%%%%%%%%%%%%%%%%%%%%%%%%%%%%%%%%%%%%%%%%%%%%%%%%%%%%%%%%%%%%%%%%%%%%%%%%%%%%%%%%%%%%%%
\end{document}